\documentclass[twocolumn]{aastex631}

\usepackage{graphicx}
\usepackage{dcolumn}
\usepackage{bm}
\usepackage{xcolor}

\begin{document}

\title{Relativistic Alfv\'en turbulence at kinetic scales}

\correspondingauthor{Cristian Vega}
\email{csvega@wisc.edu}

\author{Cristian Vega}
\affiliation{Department of Physics, University of Wisconsin at Madison, Madison, Wisconsin 53706, USA}

\author[0000-0001-6252-5169]{Stanislav Boldyrev}
\affiliation{Department of Physics, University of Wisconsin at Madison, Madison, Wisconsin 53706, USA}
\affiliation{Center for Space Plasma Physics, Space Science Institute, Boulder, Colorado 80301, USA}

\author[0000-0003-1745-7587]{Vadim Roytershteyn}
\affiliation{Center for Space Plasma Physics, Space Science Institute, Boulder, Colorado 80301, USA}



\begin{abstract}
In a strongly magnetized, magnetically dominated relativistic plasma, Alfv\'enic turbulence can extend to scales much smaller than the particle inertial scales. It leads to an energy cascade somewhat analogous to inertial- or kinetic-Alfv\'en turbulent cascades existing in non-relativistic space and astrophysical plasmas. Based on phenomenological modeling and particle-in-cell numerical simulations, we propose that the energy spectrum of such relativistic kinetic-scale Alfv\'enic turbulence is close to $k^{-3}$ or slightly steeper than that due to intermittency corrections or Landau damping. We note the analogy of this spectrum with the Kraichnan spectrum corresponding to the enstrophy cascade in 2D incompressible fluid turbulence. Such turbulence strongly energizes particles in the direction parallel to the background magnetic field, leading to nearly one-dimensional particle momentum distributions. We find that these distributions have universal log-normal statistics.   
\end{abstract}



\section{Introduction}
Large-scale astrophysical flows are often hydrodynamically unstable, which leads to the excitation of velocity, density, and electromagnetic fluctuations spanning a broad range of spatial and temporal scales. A significant fraction of energy associated with large-scale motions can then be converted into random collective plasma fluctuations and eventually dissipated at significantly smaller scales due to non-ideal processes, such as particle collisions or wave-particle interactions. In a weakly collisional plasma, the dissipated turbulent energy is converted into heat or non-thermally accelerated particles. The particle distribution functions can significantly deviate from a Maxwellian, which affects plasma dynamics and thermodynamics, as well as the radiative signatures of astrophysical objects \cite[e.g.,][]{drake2013,sironi2014,zhdankin2017a,comisso2019,zhdankin2020,nattila2020,demidem2020,trotta2020,pezzi2022,vega2022b,lemoine2022,comisso2022,vega2023}. 

Since astrophysical plasmas are typically magnetized, large-scale fluctuations are often dominated by low-frequency shear-Alfv{\'e}n modes. The energy dissipation occurs at much smaller, kinetic scales that are comparable to plasma microscales such as particle inertial and gyro scales. At such scales, the shear-Alfv{\'e}n modes transform into kinetic-Alfv{\'e}n or inertial-Alfv\'en modes. Energy cascade in the kinetic range arguably governs the energy dissipation and particle heating in a collisionless turbulent plasma, and it may be relevant for non-thermal particle acceleration. Kinetic-range modes play an important role in space and astrophysical plasmas where they have been studied analytically, numerically, and where possible, in in situ spacecraft measurements \cite[e.g.,][]{chen2013a,sahraoui2013,chen2014,told2015,he2020,chen_boldyrev2017,roytershteyn19,muni2023,vega2023,mallet2023}.   

Relativistic plasma turbulence has several features that are qualitatively different from the non-relativistic case. First, if the plasma temperature is relativistic, the Alfv\'en speed as well as plasma frequency and gyrofrequency depend on temperature, since instead of the rest-mass density, their expressions include particle energy. Second, in such a plasma the speed of sound is comparable to the speed of light. Even when the magnetic energy exceeds the relativistic kinetic energy of the particles, which is a situation somewhat analogous to a non-relativistic low-beta case, thermal corrections to the wave dispersion relation remain significant \cite[e.g.,][]{tenbarge2021,vega2022b}. This complicates the analytical consideration of relativistic turbulence. 

It has been observed in numerical simulations of strongly magnetized relativistic plasma that the  Alfv\'enic turbulence cascade indeed continues to the scales smaller than the particle inertial scales \cite[e.g.,][]{zhdankin2018c,comisso2019,vega2022b}. However, as such scales are typically not well resolved in studies of large-scale Alfv\'enic turbulence, the measurements of the energy spectrum and other statistical properties of kinetic-scale fluctuations remain inconclusive. In this work, we study numerically and analytically kinetic-range Alfv\'enic turbulence in ultra-relativistic electron-positron plasma immersed in a strong background magnetic field, $B_0\gg \delta B_0$. We discuss the spectrum of magnetic and electric fluctuations, as well as the intermittency properties of turbulence. As relativistic turbulence leads to efficient particle acceleration, we also discuss the resulting non-thermal distribution function of the accelerated particles. 

In what follows, we consider a pair plasma and reserve the standard notations, $d_e=c/\omega_{pe}$, $\omega_{pe}=\sqrt{4\pi n_0 e^2/m_e}$, and $\Omega_{ce}=|e|B_0/m_e c$ for the non-relativistic electron inertial scale, electron plasma frequency, and electron cyclotron frequency, correspondingly. Their relativistic counterparts are not unique but rather depend on the wave mode considered. Their corresponding definitions will be given in the text.

\section{Analytical consideration}
\label{section2}
Consider a relativistic electron-positron plasma with an imposed uniform background magnetic field ${\bm B}_0=B_0 {\hat {\bm z}}$. The plasma magnetization with respect to the background field is defined as 
\begin{eqnarray}
\label{sigma0}
\sigma =\frac{B_0^2}{4\pi n_0 w m_e c^2},
\end{eqnarray}
where $n_0$ denotes the unperturbed number density of electron or positron species, and $n_0 w m_e c^2$ is the corresponding enthalpy density. Assuming that the plasma particle distribution is an isotropic Maxwell-J\"uttner function with temperature $T$, the enthalpy is calculated as  
$w=K_3(1/\theta)/K_2(1/\theta)$, 
where $K_\nu $ is the modified Bessel function of the second kind.  In this formula, $\theta=k_BT/m_e c^2$ is the normalized temperature. 
Similarly, one can define the magnetization with respect to the fluctuating part of the field,
\begin{eqnarray}
{\tilde \sigma}=\frac{(\delta B)^2}{4\pi n_0 w m_e c^2}.
\label{sigma_tilde}
\end{eqnarray}

In this work, we assume that plasma is magnetically dominated, $\sigma\gg 1$, and the magnetic fluctuations are smaller than the background field, ${\tilde \sigma}\ll \sigma$. In our numerical simulations, we initialize the runs with magnetic fluctuations satisfying ${\tilde \sigma}_0\gg 1$. However, as turbulence evolves, turbulent fluctuations efficiently heat the plasma, so plasma temperature becomes ultrarelativistic while simultaneously, ${\tilde \sigma}$ decreases. This reflects the fact that relativistic turbulent motion is inherently compressible, which allows colliding fluid elements to convert their kinetic energy into heat rapidly \cite[][]{zhdankin2018c,nattila2020,vega2022a,vega2023}. We will therefore analyze the case when plasma bulk fluctuations are non-relativistic (or mildly relativistic), while plasma temperature is ultrarelativistic.\footnote{While the idealized assumption of the ultrarelativistic equation of state simplifies the formulae, it is not crucial for our analytic discussion. For mildly relativistic particle distributions, $\theta\sim 1$, such as, for instance, those obtained as pair-creation-annihilation equilibria \cite[][]{svensson1982}, one needs to add the rest-mass term to the normalized enthalpy, $w\to  w+1$, in the expression for the relativistic inertial scale $d_{rel}$, and use the general equation of state in the expression for the acoustic velocity,~$v_s$.}

Recently, a two-fluid analytic model was proposed to describe Alfv\'enic turbulence in a magnetically dominated relativistic plasma \cite[][]{vega2022b}. This model assumes that the fluctuations are anisotropic in the Fourier space with respect to the background magnetic field, $k_z\ll k_\perp$, and the magnetic and electric fluctuations are dominated by their field-perpendicular components. The equations require a closure for the plasma pressure tensor. There is no rigorous prescription for such a closure in a collisionless plasma.  In the magnetically dominated case considered here, only the field-parallel component of the pressure tensor contributes.  We may then introduce the acoustic velocity as $v_s^2=c^2(\partial P_\|/\partial u)\vert_0$, where $u$ is the thermal energy density. (For relativistic plasma temperatures, the acoustic velocity is on the order of the speed of light. For example, in the case of the 3D ultrarelativistic isotropic adiabatic equation of state, one has $P\approx u/3$ and $v_s^2\approx c^2/3$.)

The model equations govern Alfv\'enic turbulence in an ultra-relativistic $(\theta\gg 1)$ magnetically dominated ($\sigma\gg 1$) pair plasma. It is convenient to normalize the electric potential and the $z$-component of the magnetic vector potential as ${\tilde \phi}={\phi} c/B_0$ and ${\tilde A}_z={A}_z c/(B_0\sqrt{1+2/\sigma})$. The perpendicular components of the magnetic and velocity fluctuations are defined as $\delta {\tilde{\bm b}}_\perp=-{\hat {\bm z}}\times \bm{\nabla} {\tilde A}_z$ and $\delta\tilde{{\bm v}}_\perp={\hat{\bm z}}\times \bm{\nabla} {\tilde \phi}$, and they both have the dimensions of velocity. Below we will use only these variables and omit the overtilde sign. The equations then have the form \cite[][]{vega2022b}:
\begin{eqnarray}
\frac{\partial}{\partial t}\nabla^2_\perp\phi 
+\left(\hat{{z}}\times \bm{\nabla}_{\perp} \phi \right)\cdot \bm{\nabla}_{\perp} \nabla^2_\perp\phi 
=-v_A\nabla_{\|} \nabla_{\perp}^{2} A_z, \quad \label{low_beta}\\
 \frac{\partial}{\partial t} \left( A_z -d_{rel}^2\nabla^2_\perp A_z\right) - \left(\hat{{z}}\times \bm{\nabla}_\perp\phi\right)\cdot\bm{\nabla}_\perp d_{rel}^2\nabla^2_\perp {A_z} \quad \nonumber \\
 = - v_A \nabla_\| \left( {\phi} -\frac{v_s^2}{c^2}d_{rel}^2\nabla_\perp^2\phi \right),\quad
\label{reduced} 
\end{eqnarray}
where 
\begin{eqnarray}
v_A=\frac{c}{\sqrt{1+2/\sigma}}\approx c
\end{eqnarray}
is the relativistic Alfv\'en speed in a pair plasma, 
\begin{eqnarray}
d^2_{rel}=\frac{wc^2}{2\omega_{pe}^2}=\frac{wm_ec^2}{8\pi n_0e^2}    
\end{eqnarray}
is the relativistic inertial scale, and the magnetic-field-parallel gradient is given by
\begin{eqnarray}
\nabla_\|=\partial/\partial z-\frac{1}{v_A}({\hat z}\times \bm{\nabla}_\perp A_z)\cdot \bm{\nabla}_\perp .
\end{eqnarray}
The extra factor of 2 in the expressions for $v_A$ and $d_{rel}$ reflects the fact that the total plasma density is twice the electron or positron density. The very last term in the right-hand side of Eq.~(\ref{reduced}), proportional to $v_s^2/c^2$, arises from the hydrodynamic closure for the parallel-pressure term discussed above.   

In the linear case, these equations describe the waves with the dispersion relation \cite[][]{vega2022b}:
\begin{eqnarray}
\omega^2=k_z^2v_A^2\,\frac{1+\frac{v_s^2}{c^2}k_\perp^2d_{rel}^2}{1+k_\perp^2d_{rel}^2}.
\label{iaw}
\end{eqnarray}
Similarly to the kinetic-Alfv\'en waves, the numerator of this expression involves the contribution of the thermal effects. Similarly to the inertial Alfv\'en waves, the denominator includes the contribution of the electron (and positron) inertia. This is somewhat analogous to the inertial kinetic-Alfv\'en modes in a non-relativistic plasma \cite[e.g.,][]{streltsov95,lysak96,chen_boldyrev2017,roytershteyn19,loureiro2018,boldyrev2021}. In contrast with the non-relativistic case, however, in a relativistic plasma, we have $v_s\sim c$, so that the thermal contribution in the numerator is never negligible.  Rather, the thermal and inertial effects in Eq.~(\ref{iaw}) are necessarily of the same order. Moreover, since the speed of sound is comparable to the thermal speed, the Landau damping of the linear modes is also generally strong at $k_\perp^2 d^2_{rel}\gtrsim 1$. As discussed in the Appendix, the applicability of the linear dispersion relation in Eq.~(\ref{iaw}) at such scales depends on the particle distribution function. Such a function may, in general, be strongly non-thermal.


At large hydrodynamic scales, $k_\perp^2 d^2_{rel}\ll 1$, the thermal and inertial effects are negligible, and Eqs.~(\ref{low_beta}, \ref{reduced}) transform into the reduced magnetohydrodynamic equations mathematically identical to the non-relativistic case \cite[][]{tenbarge2021,vega2022b}. Available fluid and first-principle particle-in-cell kinetic simulations of relativistic turbulence at such scales \cite[e.g.,][]{zrake2012,zhdankin2018c,chernoglazov2021,vega2022b} indeed produce the energy spectra consistent with the spectrum of non-relativistic Alfv\'en turbulence \cite[e.g.,][]{boldyrev2006,boldyrev2009,mason2006,mason2012,perez_etal2012,tobias2013,chandran_intermittency_2015,chen2016,kasper2021}.

The spectrum of turbulence at small, kinetic scales, $k_\perp^2d_{rel}^2\gg 1$, is currently not well understood. The description of relativistic turbulence at such scales faces both numerical and analytical challenges. On the numerical side, available particle-in-cell kinetic simulations do not typically have a strong enough guide field $B_0$ in order to separate the inertial and gyro scales, or large enough numerical resolution (number of cells and particles) in order to reliably measure the spectra in the sub-inertial interval $1/d_{rel}^2\ll k_\perp^2 \ll 1/\rho_e^2$. The reported electromagnetic spectra at the sub-inertial scales, 
\begin{eqnarray}
W_k\,dk_\perp=\left(|E_k|^2+|B_k|^2\right)\,2\pi k_\perp dk_\perp, 
\end{eqnarray}
varied depending on the strength of the guide field. The spectrum was consistent with $W_k\propto k^{-4.5}$ for relatively weak guide fields, $B_0/\delta B_0\lesssim 1$, and flattened to approximately $W_k\propto k^{-3.5}$ for stronger fields, $B_0/\delta B_0\sim 3$ \cite[e.g.,][]{zhdankin2018c,comisso2018,nattila2020,vega2022b,vega2022a}. We will be interested in the limit of strong guide field, $B_0/\delta B_0\gg 1$; our numerical simulations discussed below will correspond to $B_0/\delta B_0=10$ and $\sigma_0=4000$.

On the analytical side, at sub-inertial scales, $k_\perp^2d_e^2\gg 1$, the model equations (\ref{low_beta}, \ref{reduced}) formally conserve the energy integral \cite[][]{vega2022b}, 
\begin{eqnarray}
{\cal E}=\frac{B^2_0 d_{rel}^2}{8\pi v_A^2}\int\left[ \left(\nabla_\perp^2 A_z\right)^2 
 +\frac{v_s^2}{c^2}\left(\nabla_\perp^2\phi\right)^2 \right]d^3 x. \qquad
\label{energy_kin}
\end{eqnarray}
This quantity is expected to cascade in a Fourier space toward large wave numbers, somewhat analogously to the enstrophy cascade in 2D incompressible hydrodynamic turbulence \cite[][]{kraichnan1967}. Dimensional arguments then predict the electromagnetic energy spectrum
\begin{eqnarray}
\label{relkaw}
W_k\,dk_\perp\propto k_\perp^{-3}\,dk_\perp.
\end{eqnarray}
Similarly to the hydrodynamic case, the energy cascade is expected to have intermittency corrections that lead to a steeper energy spectrum,
\begin{eqnarray}
\label{relkaw_int}
W_k\,dk_\perp\propto k_\perp^{-3}\ln^{-1/3}(k_\perp/k_0)\,dk_\perp.
\end{eqnarray}
Here, $k_0$ is the large-scale boundary of the spectrum, which approximately corresponds to the inverse electron inertial scale. The intermittency correction reflects the non-locality of turbulence. The electromagnetic energy spectrum close to $k_\perp^{-3}$ implies that vorticity and current structures at scales $k_\perp \gg k_0$ are strained most efficiently by turbulent eddies at scales $k_0$ \cite[e.g.][]{boffetta2012}. 

As discussed above, significant Landau damping may affect relativistic Alfv\'en turbulence at kinetic scales. We, however, conjecture that as a consequence of the non-locality of turbulence, the spectrum should exhibit a near power-law behavior, close to that given by Eqs.~(\ref{relkaw}, \ref{relkaw}).  Indeed, the Kraichnan spectrum is established due to gradient-stretching of small-scale structures by turbulent eddies of the scale~$k_0\sim 1/d_{rel}$. As a result, all small-scale modes have the same evolution time and the same parallel phase velocity. A particle resonating with structures of scales~$k_\perp \gg 1/d_{rel}$ then essentially resonates with an entire eddy of scale~$k_0\sim 1/d_{rel}$. Landau damping may therefore regulate the overall intensity of kinetic-scale fluctuations, while not significantly affecting their spectrum. Our numerical results analyzed below seem to be consistent with this prediction.


Finally, we mention a formal restriction on a turbulent cascade in a magnetically dominated plasma, which applies in both relativistic and non-relativistic cases. When the magnetization parameter $\sigma_0$ is large,\footnote{For non-relativistic temperatures, the magnetization parameter~(\ref{sigma0}) becomes the so-called plasma quasineutrality parameter, $\sigma_0=\Omega^2_{ce}/\omega^2_{pe}$ \cite[e.g.,][]{vega2022a,vega2022b}.} magnetic fluctuations at small enough scale $\lambda_s$ may enter the so-called  ``charge-starved" regime when the scale satisfies \cite[e.g.,][]{thompson2006,thompson2008,blaes1990,boldyrev2021,chen2022,nattila2022}:
\begin{eqnarray}
(\delta B_{\lambda_s}/B_0)^2(d_e/\lambda_s)^2\sim 1/\sigma_0 .   
\end{eqnarray}
At such scales, the field-parallel electric current would correspond to the electrons moving with the speed of light.  In this regime, the non-relativistic equation~(\ref{reduced}) for parallel electron dynamics cannot be used. 
Since the current is restricted by the speed of light, its autocorrelation function is limited by a constant, 
\begin{eqnarray}
\langle J_z({\bf x})J_z({\bf x}')\rangle < 4n_0^2 q^2 c^2. 
\end{eqnarray}
We then conclude that in the asymptotic limit $k\lambda_s \gg 1$, the Fourier spectrum of the current cannot be flatter than~$k_\perp^{-1}$ and, correspondingly, the spectrum of the magnetic field cannot be flatter than~$k_\perp^{-3}$. The spectra given by Eqs.~(\ref{relkaw}, \ref{relkaw_int}) satisfy this restriction.


\section{Numerical results}

To address this problem numerically, we performed two 2.5D particle-in-cell simulations of decaying turbulence in a pair plasma with the fully relativistic code VPIC \cite[][]{bowers2008}. In 2.5D simulations, vector fields have three components but depend on only two spatial coordinates $x$ and $y$. Similarly, the particle distribution function depends on three velocity components. A uniform background magnetic field is imposed in the $z$-direction, ${\bm B}_0=B_0{\hat z}$. The simplified ``$2.5$''-dimensional setup allows us to afford a relatively high numerical resolution of kinetic-range turbulence. Since all the vector components of the electromagnetic field and particle momenta are preserved, it is expected to capture some essential nonlinear interactions existing in the 3D case. Previous numerical studies involving 2.5D and 3D runs seem to produce similar energy spectra of fields and particles \cite[e.g.,][]{zhdankin2017a,zhdankin2018c,comisso2018,comisso2019,vega2023}.

\begin{figure*}[]
\centering
\includegraphics[width=0.95\columnwidth]{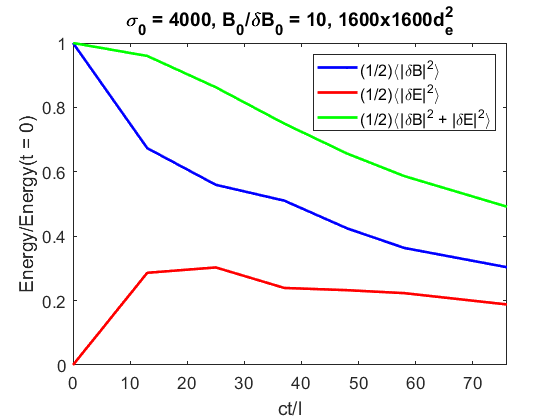}
\includegraphics[width=0.95\columnwidth]{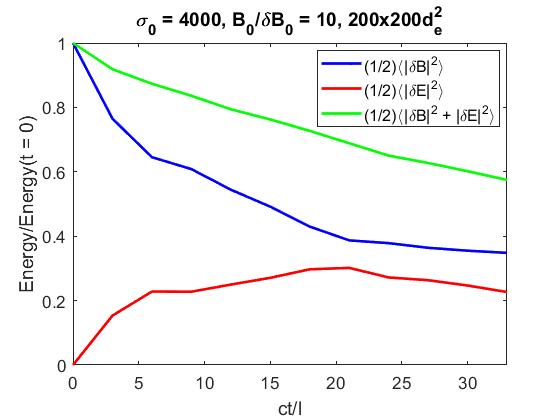}
\caption{Time evolution of the electromagnetic energy.}
\label{time_energy}
\end{figure*}

\begin{figure*}[]
\centering
\includegraphics[width=\columnwidth]{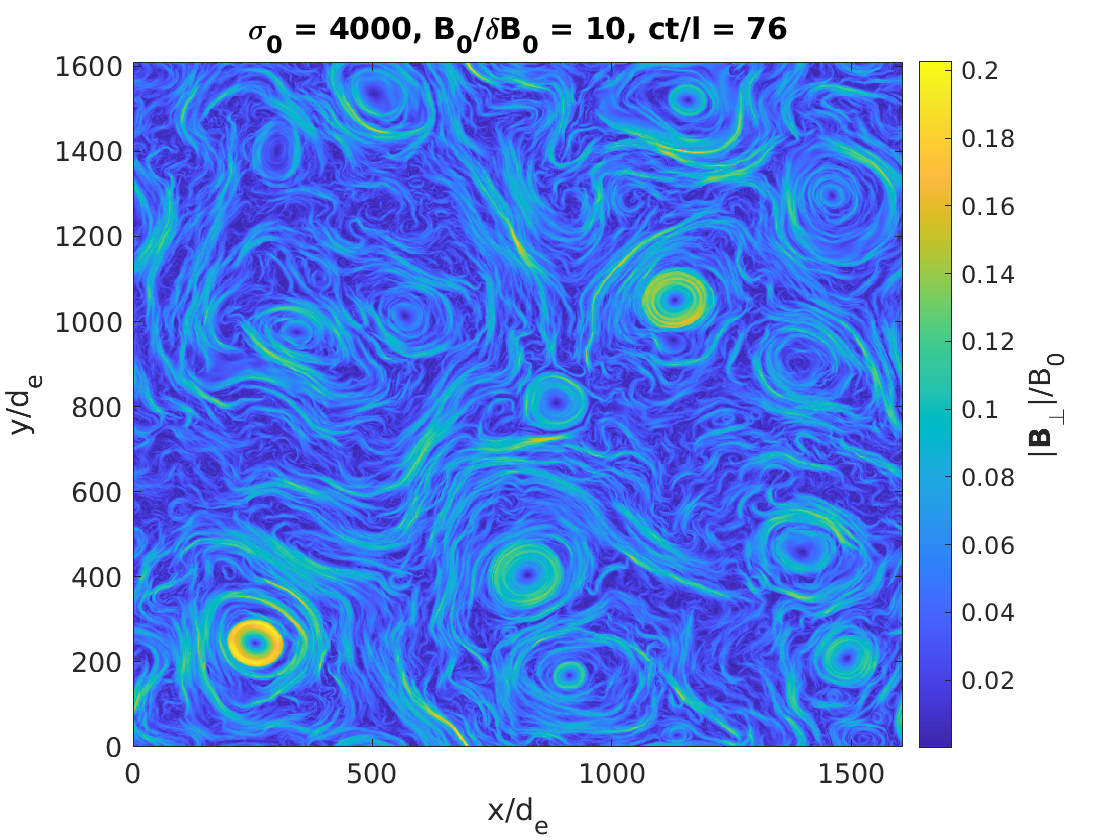}
\includegraphics[width=\columnwidth]{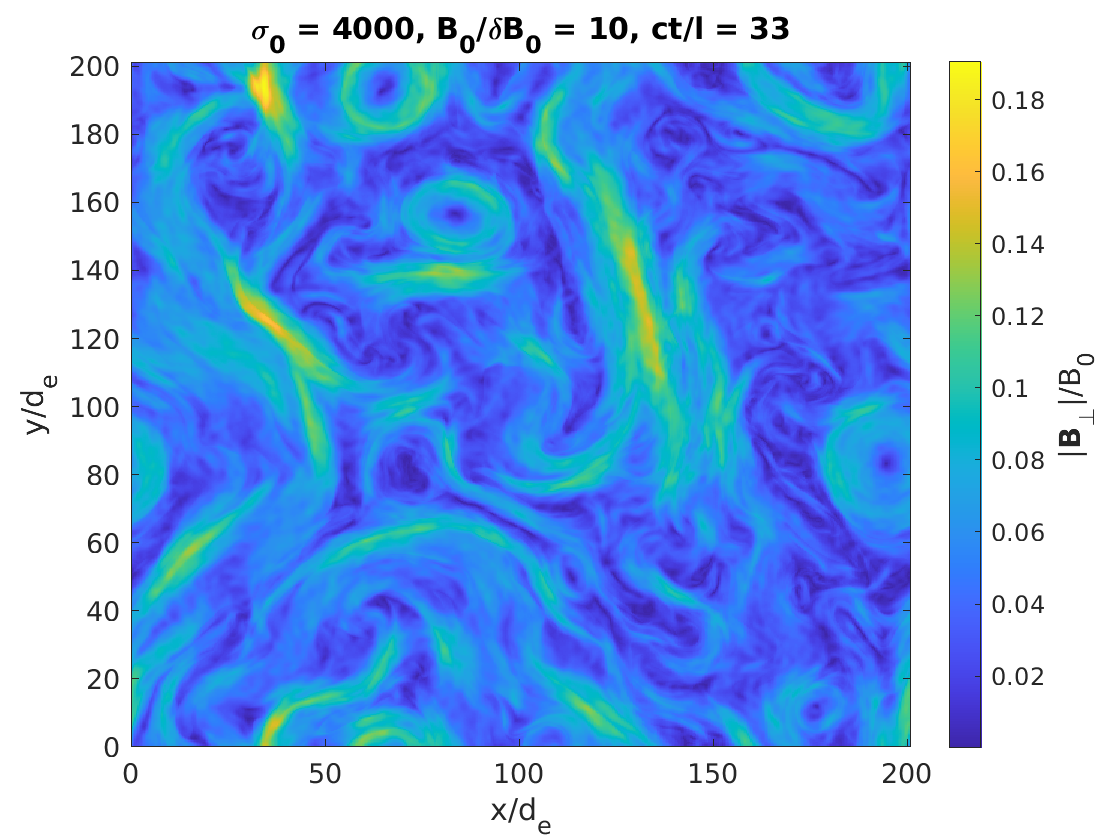}
\caption{Visualization of magnetic fluctuations shows similar structures formed in large- and small-box runs.}
\label{structures}
\end{figure*}

Table \ref{table} summarizes the parameters of the simulations. Run I is a large-box, high-resolution simulation that spans both hydrodynamic and kinetic scales, with about 15 cells per nonrelativistic $d_e$. Run II is a small-box simulation where the number of cells per nonrelativistic $d_e$ was increased to 80, drastically improving the resolution of the sub-$d_e$ fluctuations while decreasing the hydrodynamic range. Both simulation domains are double periodic $L\times L$ squares with 100 particles per cell per species.

\begin{table}[h!]
\centering
\begin{tabular}{c c c c c} 
\hline
{Run} & {Size} ($d_e^2$) & {Res. (\# of cells )} & $n_{\text{max}}$ & $\omega_{pe}\delta t$ \\
\hline
I & $1600^2$ & $23552^2$ & 8 & $1.2\times10^{-2}$\\ 
II & $200^2$ & $16640^2$ & 4 & $3.5\times10^{-4}$\\ 
\hline
\end{tabular}
\caption{The parameters of the runs.}
\label{table}
\end{table}

The turbulence is seeded by randomly phased magnetic perturbations of the Alfv\'enic type
\begin{eqnarray}
\delta{\bm B}(\mathbf{x})=\sum_{\mathbf{k}}\delta B_\mathbf{k}\hat{\xi}_\mathbf{k}\cos(\mathbf{k}\cdot\mathbf{x}+\phi_\mathbf{k}),
\end{eqnarray}
where the unit polarization vectors are normal to the background field, $\hat{\xi}_\mathbf{k}=\mathbf{k}\times {\bm B}_0/|\mathbf{k}\times{\bm B}_0|$. The wave vectors of the modes are given by $\mathbf{k}=\{2\pi n_x/L,{2}\pi n_y/L\}$, where $n_x,n_y=1,...,n_{\text{max}}$. All modes have the same amplitudes $\delta B_{\mathbf{k}}$. The time is normalized to the large-scale crossing time $l/c$, where $c$ is the speed of light and the outer scale of turbulence is evaluated as 
$l=L/n_{\text{max}}$,
with $n_{\text{max}}=4$~or~$8$ depending on the considered run, see Table~\ref{table}.

The initial distribution of plasma particles is an isotropic Maxwell-J\"uttner distribution with the temperature parameter $\theta_0=0.3$. Here, $\theta_0=k_BT_0/m_ec^2$ is the normalized initial temperature. For $\theta_0=0.3$, one has $w_0\approx 1.88$, for ultrarelativistic temperatures, $\theta_0\gg 1$, one has $w_0\approx 4\theta_0$, while for non-relativistic plasma, $\theta_0\ll 1$,  $w_0\approx 1$. In the large-box run~I, an initial plasma current is also added to the system with the goal to compensate for the curl of the initial magnetic perturbations, $J_z=(c/4\pi){\bm \nabla}\times \delta {\bm B}_0$. This helps to avoid the generation of high-frequency ordinary modes with non-zero $E_z$ in addition to the low-frequency Alfv\'en modes.  {To add the current, the initial plasma density $n_0$ is kept uniform, and velocity $U_{z}^s = J_z/(2 q_s n_0)$ is added (in a Newtonian way}) to each particle of species $s$ with charge $q_s = \pm e$ (positrons and electrons) sampled from the Maxwell-J\"uttner distribution, provided $|v_z^s+U_{z}^s| < c$. The distribution is unchanged in the regions where such a velocity increase would lead to $|v_z^s+U_{z}^s| \geq c$. The addition of such a current does not change the core of the particle distribution function but modifies its high-energy tail, as will be seen below. 

\begin{figure*}[]
\centering
\includegraphics[width=0.97\columnwidth]{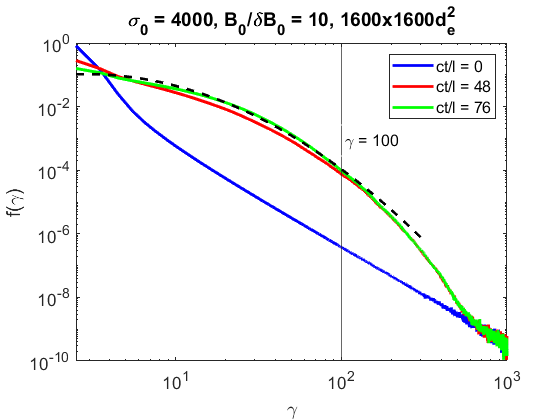}
\includegraphics[width=0.97\columnwidth]{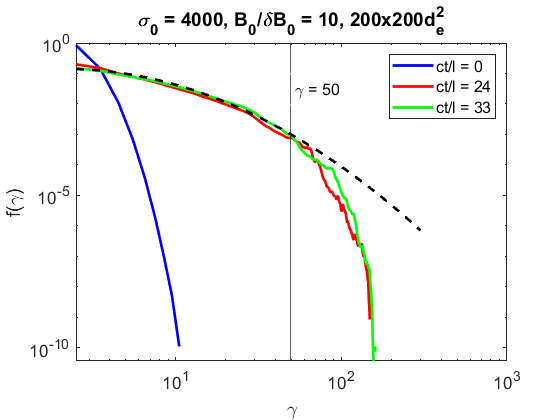}
\caption{Electron energy distribution functions. The black dashed lines show the log-normal fits with the parameters given in Table~\ref{table2}. The thin vertical lines show the ranges of $\gamma$ where the fits are made.}
\label{particles1}
\end{figure*}

\begin{figure*}[]
\centering
\includegraphics[width=\columnwidth]{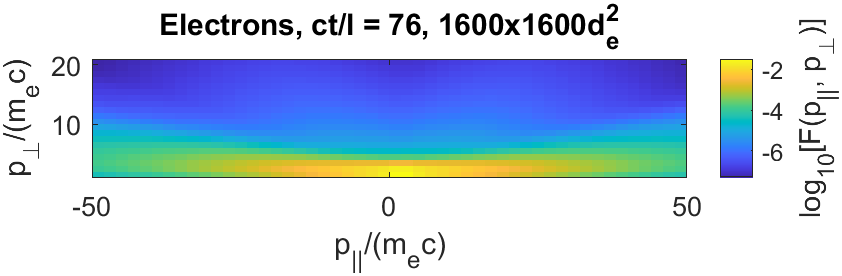}
\includegraphics[width=\columnwidth]{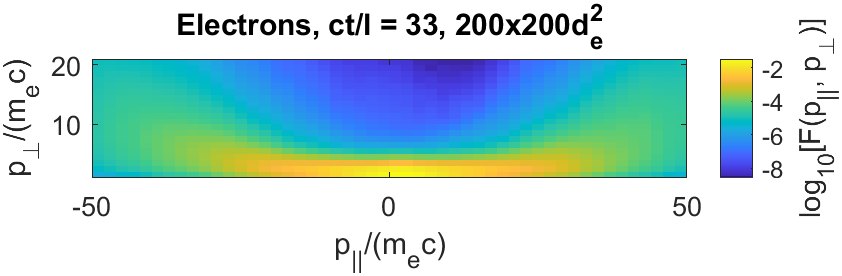}
\caption{The electron distribution functions. Here, $p_\|$ and $p_\perp$ are particle momenta in the directions parallel and perpendicular to the background magnetic field ${\bm B}_0$. The functions are strongly anisotropic. The anisotropy at very large energies is slightly less pronounced in run II, which is likely a consequence of the smaller box size.}
\label{particles2}
\end{figure*}

In the small-box run~II, the addition of a compensating current is less practical, since it would formally require the electron velocities to exceed the speed of light in more cells. 
We do observe the generation of a weak ordinary mode in this case. The presence or absence of the initial compensating current, however, does not qualitatively affect the particle distribution function and the spectra of Alfv\'enic fluctuations eventually generated by the developed turbulence.   

We define the two initial magnetization parameters $\sigma_0$ and $\tilde{\sigma}_0$ by using Eqs.~(\ref{sigma0})~and~(\ref{sigma_tilde}), where we substitute the values of the initial root-mean-square magnetic fluctuations, $\delta B_{0}=\langle\delta B^2({\bf x}, t=0)\rangle^{1/2}$, and the normalized initial enthalpy per particle, $w_0=K_3(1/\theta_0)/K_2(1/\theta_0)$. These magnetization parameters in our simulations are $\sigma_0=4000$ and $\tilde{\sigma}_0=40$ (i.e., $\delta B_0/B_0=1/10$).


Figure~\ref{time_energy} illustrates the time evolution of the electromagnetic energies in both runs. Our statistical analysis of field and particle distribution functions is performed after the initial relaxation is completed, that is, after quasi-steady states of electric and magnetic fluctuations are reached. Figure~\ref{structures} presents a visualization of magnetic fluctuations in both runs, which shows similar morphologies of the magnetic structures generated in the large- and small-box simulations.

In both runs, turbulent fluctuations lead to significant particle energization.  The particle energy distribution functions evolve fast until about half of the initial magnetic energy has been converted into the kinetic energy of the particles. After that, the distributions reach quasi-saturated states, as shown in Figure~\ref{particles1}. In agreement with previous studies \cite[][]{nattila2022,vega2022a}, the particle momentum distribution functions are strongly non-thermal (non-Maxwell-J\"uttner). Moreover, they are strongly anisotropic with respect to the background magnetic field, see Figure~\ref{particles2}. The electrons are energized mostly in the field-parallel direction, leading to a nearly one-dimensional particle distribution function.   

We find that for $\gamma>1$, the particle energy distribution functions can be well approximated by the log-normal distribution:
\begin{eqnarray}
f(\gamma)\,d\gamma=\frac{A}{\gamma }\exp\left(-\frac{\left(\ln\gamma-\mu\right)^2}{2\sigma_s^2} \right)d\gamma,
\label{lnfit}
\end{eqnarray}
where $A$ is a normalization constant. Here, we use the fact that the particle energy can be represented as $\gamma mc^2$, so that the relativistic energy distribution can be characterized by the distribution of the corresponding Lorenz factors~$\gamma$. The parameters $\mu$ and $\sigma_s^2$ of the log-normal fits for each run are given in Table~\ref{table2}, which also shows the measured average Lorenz factors of the electron energies.\footnote{We note that both the plasma magnetization and the statistical standard deviation are conventionally denoted by the same letter~$\sigma$. We, therefore, denote the standard deviation by $\sigma_s$; it should not be confused with the plasma magnetization parameter.} To find the best fits, we varied the parameters  $\mu$ and $\sigma_s^2$ with an imposed constraint that the average Lorenz factors, $\langle\gamma\rangle=\exp\left(\mu+\sigma_s^2/2\right)$, have the fixed values presented in the Table. Table~\ref{table2} also shows the ratio of the field-parallel and field-perpendicular components of the electron pressure tensor. Since the bulk plasma motion is only mildly relativistic, the ratio of the electron pressure tensor components is numerically calculated as the ratio of the components of the electron energy-momentum tensor, $P_\perp/P_\| \approx T_\perp/T_{zz}$. 

We also note that the {\em log-normal} particle energy distributions generated by turbulence with a strong guide field are different from {\em power-law} distributions numerically observed in turbulence with a weak guide field \cite[e.g.,][]{zhdankin2017a,zhdankin2018c,comisso2019,vega2022a}. This may indicate that different particle acceleration mechanisms operate in these two regimes. In the case of a strong guide field, a particle gyroradius is much smaller than the plasma inertial scale. In this limit, the acceleration may be provided by the electric field parallel to the background magnetic field or by the particle curvature drifts in the vicinity of strong magnetic structures. As a result, particles are accelerated in the field-parallel direction. Particles with sufficiently high energies can however have gyroradii larger than the plasma inertial scale \cite[e.g.,][]{vega2022a}, which happens when $\gamma c/\Omega_{ce}>d_{rel}$, that is,  
\begin{eqnarray}
\gamma > \gamma_{crit}=\sqrt{\frac{w w_0 \sigma_0}{2}}. 
\end{eqnarray}
Such particles interact with Alfv\'enic turbulent fluctuations more efficiently. Their magnetic moments may be not conserved, and they can be accelerated in the field-perpendicular direction. As a result, their energy distribution functions develop power-law tails. Since in our simulations, $w_0\approx 1.88$, $\sigma_0=4000$, and $w\approx 2\langle \gamma\rangle$ (see Eq.~(\ref{w1D})), we obtain a rather large value of the critical energy, $\gamma_{crit}\approx 274$.
This may explain why we observe only the log-normal part of the distributions.


As the initial field-perpendicular magnetic perturbations relax, they drive turbulence at large scales. In a magnetically dominated plasma, the excited large-scale fluctuations can be a combination of the two modes whose magnetic polarizations are normal to the background field: the shear-Alfv\'en mode and the ordinary mode. The frequency of the ordinary mode is given by
\begin{eqnarray}
\omega^2=\omega_{p}^2+k^2c^2,
\label{o_mode}
\end{eqnarray}
where $\omega^2_p=2\omega^2_{pe}\langle1/\gamma^3 \rangle$, see Eqs.~(\ref{rel_omega}) and~(\ref{omega_o}) in Appendix.  This mode is excited in our setup with a relatively low amplitude, contributing only a small fraction of the total turbulent energy.  Such a mode is not important at the hydrodynamic scales. 

\begin{figure*}[]
\centering
\includegraphics[width=\columnwidth]{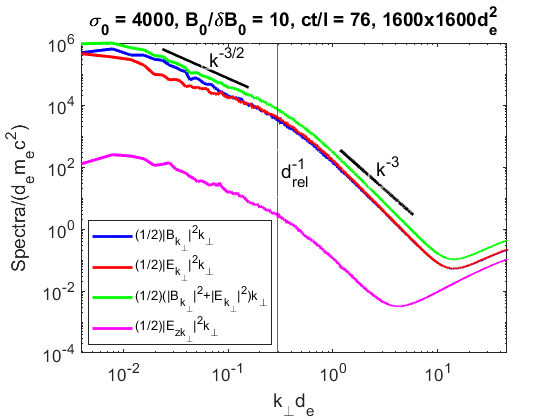}
\includegraphics[width=\columnwidth]{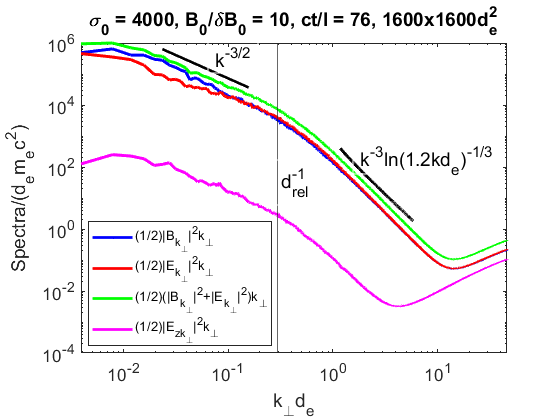}
\caption{Spectra of electric and magnetic fluctuations in the large-box run~I. The total energy spectrum is slightly steeper than $k^{-3}$, but is consistent with the spectrum including a logarithmic intermittency correction, cf.~Eqs~(\ref{relkaw})~and~(\ref{relkaw_int}). }
\label{EM_large}
\end{figure*}

The energies of electric and magnetic fluctuations in the large-box run are shown in Fig.~\ref{EM_large}. The total energy spectrum is slightly steeper than $k^{-3}$, however, it is consistent with the Kraichnan spectrum of turbulence including a logarithmic intermittency correction, in agreement with~Eq.~(\ref{relkaw_int}). We find that $k_0^{-1}=1.2\, d_e$ provides a good match. Also, as we mentioned before, Landau damping may play a role in the steepening of the spectrum. 

\begin{table}[h!]
\hskip-2.0cm
\centering
\begin{tabular}{ c c c c c} 
\hline
{\small Run} & $\mu$ & $\sigma_s^2$ & $\langle\gamma\rangle$ & $P_\perp/P_\|$\\
\hline
I &  1.97 & 0.92 &  11.3 & 0.017\\ 
II & 1.75 & 1.01 & 9.5 & 0.023 \\
\hline
\end{tabular}
\caption{Parameters of the particle distribution functions. The measurements of the coefficients of the log-normal fit in Eq.~(\ref{lnfit}), the average electron Lorentz factor $\langle\gamma\rangle$, as well as the components of the electron pressure tensor, are made at $ct/l = 76$ for run I and $ct/l = 33$ for run II.}
\label{table2}
\end{table}

In \sout{a} \textcolor{red}{the} small-box run, the imposed magnetic fluctuations at $t=0$ are relatively strong at small scales; their decay leads to the production of a weak ordinary mode. Such a mode is most strongly generated at the smallest scale available for Aflv\'enic fluctuations since the current is largest there.  This is the inertial scale of a non-relativistic pair plasma, expressed as 
\begin{eqnarray}
\label{dnonrel}
d_{nonrel}=d_e/\sqrt{2}.
\end{eqnarray}
We will see below that the energy of the ordinary modes is indeed concentrated at these kinetic scales that are the focus of our study. Since the phase velocity of the ordinary mode exceeds the speed of light, see Eq.~(\ref{o_mode}), such fluctuations are not significantly damped.    Therefore, we need to make sure that in our statistical analysis, the fluctuations associated with such a mode are separated from the Alfv\'enic fluctuations. 

\begin{figure}[]
\centering
\includegraphics[width=\columnwidth]{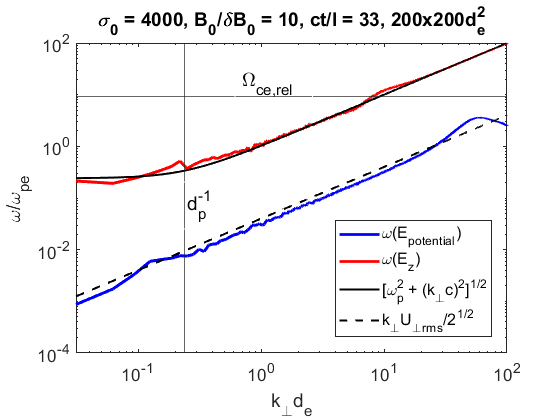}
\caption{Frequencies of the ordinary fluctuations and the potential fluctuations measured in Run~II according to Eqs.~(\ref{omega_o})~and~(\ref{omega_A}).  Here, we denote $d_p\equiv c/\omega_p$, where the relativistic plasma frequency $\omega_p$ is defined in Eq.~(\ref{o_mode}), and $\Omega_{ce, rel}\equiv \Omega_{ce}/\langle \gamma \rangle$, where $\Omega_{ce}$ is the non-relativistic electron cyclotron frequency and $\langle \gamma \rangle$ is given in Table~\ref{table2}.}
\label{spec_plus_o}
\end{figure}

The ordinary mode can be detected in simulations if we observe that its electric field is polarized in the $z$ direction, while the electric polarization of the shear-Alfv\'en mode is normal to the $z$ direction. The frequency of the $E_z$ fluctuations can be numerically found from the Maxwell-Ampere law by calculating the ratio:
\begin{eqnarray}
\omega^2=\frac{\left| ({\bm \nabla}\times {\bm B})_{z, k}-\frac{4\pi}{c}{J}_{z, k} \right|^2}{\left|\frac{1}{c} E_{z,k}\right|^2}.
\label{omega_o_num}
\end{eqnarray}
Figure~\ref{spec_plus_o} shows this frequency calculated for the small-box run. The frequency indeed agrees with the analytic dispersion relation given by~Eq.~(\ref{o_mode}). Since $k_\perp\gg k_z$, the frequency of the ordinary mode is much larger than the frequency of shear-Alfv\'en fluctuations. The oblique shear-Alfv\'en fluctuations are characterized by a mostly potential electric field. The frequency of the potential electric fluctuations can be similarly measured from the Maxwell-Ampere law:
\begin{eqnarray}
\omega^2=\frac{\left| \frac{4\pi}{c}\,{\bm k}\cdot{\bm J}_{k} \right|^2}{\left|\frac{1}{c} \,{\bm k}\cdot{\bm E}_{k}\right|^2}.
\label{omega_A}
\end{eqnarray}
Fig.~\ref{spec_plus_o} illustrates that it is indeed much smaller than the frequency of the ordinary mode. We also note that the frequency calculated in Eq.~(\ref{omega_A}) coincides with the frequency of electric charge fluctuations since the divergence of the electric field coincides with the charge density.

\begin{table}[h!]
\centering
\begin{tabular}{c c c c c c c } 
\hline
{\small Run}  &  $U^e_{z, {rms}}$ &  $U^e_{\perp, {rms}}$ & $U_{z, rms}$ & $U_{\perp, rms}$ & $\langle{\tilde \gamma^e}\rangle$ \\
\hline
I & $0.47c$ & $0.070c$ & $0.32c$ & $0.053c$ & 1.25\\ 
II & $0.61c$ & $0.079c$ & $0.39c$ & $0.056c$ & 1.73\\
\hline
\end{tabular}
\caption{Parameters of the velocity fluctuations for the electron fluid, and for the bulk plasma motion, ${\bf U}=(n_i{\bf U}^i+n_e{\bf U}^e)/(n_i+n_e)$. Here, ${\tilde \gamma^e}$ is the Lorenz factor associated with the electron fluid velocity. The measurements are made at $ct/l = 76$ for run I and $ct/l = 33$ for run II.}
\label{table3}
\end{table}

The nearly linear scaling of the measured frequency of potential fluctuations with the wavenumber can be a consequence of two independent effects. First, it may reflect the almost linear dispersion relation of the Alfv\'en mode given by Eq.~(\ref{iaw}). Indeed, in our 2D runs, the magnetic-field direction deviates from the $z$-axis by a small angle, $\sin(\theta)\sim\delta B_\perp/B\sim 0.1$. Alfv\'en fluctuations with the wavenumber $k_\perp$ then correspond to the field-parallel wavenumber $k_\|\sim k_\perp \sin(\theta)$. According to Eq.~(\ref{iaw}), this gives the frequency of linear Alfv\'en mode, $\omega\sim k_\perp c\, \sin(\theta) \sim 0.1\, k_\perp c$, which is not far away from the measurements in Fig.~\ref{spec_plus_o}. Second, it may correspond to the linear ``Doppler shift" of the frequency provided by the passive advection of the small-scale plasma structures by large-scale Alfv\'en fluctuations, $U_{\perp, rms}$.  Since $U_{\perp, rms}\sim 0.1\,c$ in our runs (see Table~\ref{table3}), the corresponding angle-averaged frequency shift, $\omega\sim k_\perp U_{\perp,rms}/\sqrt{2}$, is also consistent with the measurements in Fig.~\ref{spec_plus_o}. 

\begin{figure*}[]
\centering
\includegraphics[width=\columnwidth]{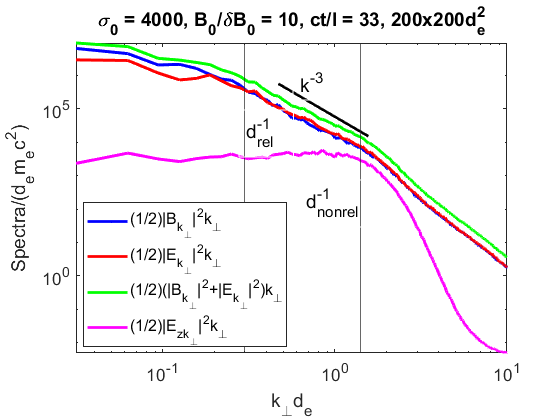}
\includegraphics[width=\columnwidth]{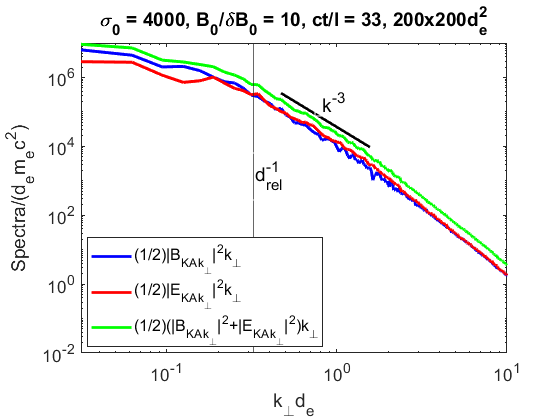}
\caption{The electric and magnetic spectra for the small-box run~II. The left panel shows the spectra of the perpendicular components of electric and magnetic fields, as well as the spectrum of $E_z$. The $E_z$-spectrum is concentrated at the non-relativistic inertial scale defined in Eq.~(\ref{dnonrel}) and corresponds to the ordinary mode produced by the initial magnetic perturbations.  The right panel shows the kinetic-Alfv\'en spectra where the fluctuations associated with the ordinary mode have been removed. The spectra indicated by the solid lines are given for the reader's orientation.}
\label{spec_clean}
\end{figure*}

The magnetic-field spectrum associated with the ordinary mode is found from the Faraday law:
\begin{eqnarray}
|{\bm B}_{k_\perp}|^2=\frac{k_\perp^2 c^2}{\omega^2}|E_{z,k_\perp}|^2,
\label{B_o}
\end{eqnarray}
where the frequency should be substituted from Eq.~(\ref{omega_o_num}). Since the frequency of the ordinary fluctuations is much larger than the frequency of the Alfv\'en mode, we may average high-frequency ordinary fluctuations independently of the low-frequency Alfv\'en fluctuations. We may then obtain the electromagnetic spectrum associated with the Alfv\'en modes by subtracting the spectrum of the ordinary mode (\ref{B_o}) from the total spectrum of magnetic fluctuations.  Fig.~\ref{spec_clean}, left panel, shows the spectra of electric and magnetic fields in the small-box run~II. The right panel shows the spectra of the electric and magnetic fields where the fluctuations associated with the ordinary mode have been removed. The observed spectrum is close to $k^{-3}$, which is consistent with the Kraichnan spectrum expected for the kinetic-scale turbulence; see Eqs.~(\ref{relkaw})~and~(\ref{relkaw_int}).

\section{Conclusions}
We presented an analytical and numerical study of kinetic-scale Alfv\'en turbulence in a magnetically dominated ultrarelativistic pair plasma. We assumed that a relatively strong background magnetic field is imposed on a plasma, which is a situation relevant, e.g., for magnetospheres of neutron stars \cite[][]{arons1986,gedalin1998,nattila2022}. This leads to the separation of the inertial and gyro scales and the existence of the kinetic interval of turbulence at scales $1/d_{rel} \ll k\ll 1/\rho_e$. Such kinetic-scale turbulence may be relevant for energy dissipation and particle energization in a turbulent plasma, and it is somewhat analogous to the kinetic-Alfv\'en or inertial-Alfv\'en turbulence previously studied in non-relativistic cases. We however demonstrated that the kinetic-scale energy cascade in the ultrarelativistic case is qualitatively different from the non-relativistic counterparts.

First, contrary to non-relativistic kinetic-Alfv\'en or inertial-Alfv\'en turbulence, the thermal and inertial effects are necessarily on the same order in the ultrarelativistic kinetic-scale turbulence. Second, the scaling of the energy spectrum is slightly steeper than $k^{-3}$, which is different from the kinetic-Alfv\'en case (the energy spectrum $\sim k^{-8/3}$ \cite[e.g.,][]{alexandrova09,boldyrev12b,tenbarge2012,boldyrev_etal2015,muni2023}) and the inertial-Alfv\'en case (the spectrum is $\sim k^{-11/3}$ \cite[e.g.,][]{loureiro2018,milanese2020}). We have proposed that in the ultrarelativistic case, the energy spectrum is consistent with the Kraichnan spectrum of incompressible 2D turbulence corresponding to the enstrophy cascade, $k^{-3}$ or $k^{-3}\ln^{-1/3}(k/k_0)$ if the intermittency corrections are taken into account. We noted that the kinetic-scale cascade may be affected by Landau damping which, in general, is not weak in the relativistic case. The spectrum, however, exhibits a near power-law behavior despite being affected by the damping. We conjectured that this might be the consequence of the non-locality of kinetic-scale turbulence.


We performed numerical simulations for two cases, with the box size much larger than the electron inertial scale $d_{rel}$ and the box size moderately larger than the inertial scale. In the first case, the hydrodynamic Alfvénic cascade ($kd_{rel}\ll 1$) was well established, while the kinetic-scale turbulence was resolved with about $15$~cells per non-relativistic inertial scale $d_e$. In the small-box case, the hydrodynamic interval was shorter, but the kinetic-scale turbulence was resolved significantly better, with $80$~cells per~$d_e$. The small box also had stronger initial magnetic fluctuations at the kinetic scales. As a result, the ``contamination" of the spectrum by the O-mode was stronger, and the measurement of the power-law exponent of the kinetic energy spectrum was less straightforward in the small-box run. In both cases, however, the spectrum of kinetic scale turbulence was consistent with the Kraichnan scaling, from $k^{-3}$ to $k^{-3}\ln^{-1/3}(k/k_0)$.
 
It is known that particles get strongly energized in collisionless magnetically-dominated relativistic turbulence \cite[e.g.,][]{zhdankin2017a,comisso2019,nattila2020,vega2022a}. In the case of a strong guide field considered here, the particles are accelerated predominantly in the direction parallel to the background magnetic field \cite[e.g.,][]{nattila2022}, leading to a quasi-one-dimensional particle distribution function.   We found that in both cases of the large and small boxes, the resulting particle distribution functions are qualitatively the same; they are well approximated by a universal log-normal distribution. This is in contrast with particle acceleration in weak guide-field turbulence, where energetic particles develop power-law energy distribution functions \cite[e.g.,][]{zhdankin2019,zhdankin2020,comisso2019,vega2022a,vega2023}, which may indicate different acceleration mechanisms in these two cases.

\begin{acknowledgments}
We thank the referee for the valuable comments. This material is based upon work supported by the U.S. Department of Energy, Office of Science, Office of Fusion Energy Sciences under award number DE-SC0024362. The work of CV and SB was also partly supported by the NSF grant PHY-2010098 and the Wisconsin Plasma Physics Laboratory (US Department of Energy Grant DE-SC0018266). VR was also partly supported by NASA grant 80NSSC21K1692. Computational resources were provided by the Texas Advanced Computing  Center (TACC) at the University of Texas at Austin and by the NASA High-End Computing (HEC) Program through the NASA Advanced Supercomputing (NAS) Division at Ames Research Center. This work also used Bridges-2 at Pittsburgh Supercomputing Center. TACC and Bridges-2 access was provided by allocation TG-ATM180015 from the Advanced Cyberinfrastructure Coordination Ecosystem: Services \& Support (ACCESS) program, which is supported by National Science Foundation grants \#2138259, \#2138286, \#2138307, \#2137603, and \#2138296.
\end{acknowledgments}

\appendix
\section{Low-frequency waves in a relativistic plasma}
\label{ap}
We consider a strongly magnetized, magnetically dominated pair plasma, where the cyclotron frequency is much larger than the plasma frequency and the frequency of the considered wave modes. Here, we imply the relativistic versions of the cyclotron and plasma frequencies that depend on the details of the particle distribution function, see the discussion below. 
In what follows, we denote by $\omega_{pe}=\sqrt{4\pi n_0 e^2/m_e}$ the non-relativistic electron plasma frequency. We also assume that the electron gyroscale is negligibly small,  $k_\perp^2\rho_{e}^2\ll 1$.  

It is convenient to choose the coordinate frame such that ${\bf k}=(k_{\perp},0,k_{z})$, where $z$ is the direction along the background magnetic field. Under these conditions, the plasma dielectric tensor simplifies to:
\begin{eqnarray}
\varepsilon_{lm}=\left(\begin{array}{c c c}
1{~~} & 0{~~} & 0\\
0{~~} & 1{~~} & 0\\
0{~~} & 0{~~} & P
\end{array}\right),
\end{eqnarray}
where the function $P(\omega, {\bm k})$ depends on the particle distribution function and will be discussed later. In order to find the frequencies and polarizations of the plasma modes, we need to solve the wave equation:
\begin{eqnarray}
\left(k^{2}\delta_{lm}-k_{l}k_{m}-\frac{\omega^{2}}{c^{2}} \varepsilon_{lm}\right)E_{m}=0,
\end{eqnarray}
which in the matrix form reads
\begin{eqnarray}
{\left(\begin{array}{ccc}
k_{z}^{2}-\frac{\omega^{2}}{c^{2}} & 0 & -k_{z}k_{\perp}\\
0 & k^{2}-\frac{\omega^{2}}{c^{2}} & 0\\
-k_{z}k_{\perp} & 0 & k_{\perp}^{2}-\frac{\omega^{2}}{c^{2}}P
\end{array}\right)}\left(\begin{array}{c}
E_{x}\\
E_{y}\\
E_{z}
\end{array}\right)=0.
\end{eqnarray}

Equating the determinant of the matrix to zero, we obtain:
\begin{eqnarray}
\left(k^{2}-\frac{\omega^{2}}{c^{2}}\right)\left(\left[k_{z}^{2}-\frac{\omega^{2}}{c^{2}}\right]\left[k_{\perp}^{2}-\frac{\omega^{2}}{c^{2}}P\right]-k_{z}^{2}k_{\perp}^{2}\right)=0.\,\,\,\quad
\end{eqnarray}
Setting the first multiplicative term to zero, one gets the dispersion relation of the electromagnetic {\it extraordinary} mode,  
\begin{eqnarray}
\omega^{2}=k^{2}c^{2},
\end{eqnarray}
whose electric-field polarization is normal to both the background magnetic field and the wave vector, 
\begin{eqnarray}
{\bm E}_{\mbox{\tiny X}}= (0, E_y, 0). 
\end{eqnarray}
By equating the second term to zero, we obtain
\begin{eqnarray}
\label{disp}
k_{\perp}^{2}+k_{z}^{2}P-\frac{\omega^{2}}{c^{2}}P=0.
\end{eqnarray}
In order to specify the function $P$ in this expression, we need to know the particle distribution function. Following \cite[][]{godfrey1975,arons1986,gedalin1998}, we assume that the particle velocity distribution function is one-dimensional, $f({\bm u})={\tilde f}(u_z)\delta({\bm u}_\perp)$, with the normalization 
\begin{eqnarray}
\int f({\bm u})\,d^3 u=\int\limits_{-\infty}^{\infty} {\tilde f}(u_z)\, du_z=1.
\end{eqnarray}
In this expression, we denote ${u}_z={v}_z/\sqrt{1-v_z^2/c^2}$, where ${v}_z$ is the particle velocity, so that $\gamma^2=1+u_z^2/c^2$. This simplifying assumption is motivated by two independent considerations. First is the fact that in a very strong guide magnetic field such as that relevant, e.g., for pulsar magnetospheres and winds, \cite[e.g.,][]{arons1986,gedalin1998}, the field-perpendicular components of particle momenta are significantly reduced with respect to their field-parallel ones due to strong synchrotron cooling. Second is the numerical observation that magnetically dominated Alfv\'enic turbulence (${\tilde \sigma}\gg 1$) strongly heats plasma particles in the field-parallel rather than field-perpendicular direction~\cite[][]{nattila2022}. We also note that in astrophysical applications, plasma can stream along the background magnetic field with relativistic velocity. Our consideration will apply to the plasma rest frame, where we assume that the particle momentum distribution is symmetric with respect to the $\pm z$-directions.   

In the considered limit of a very strong large-scale magnetic field and a one-dimensional particle velocity distribution function, one obtains for a pair plasma \cite[][]{gedalin1998}:
\begin{eqnarray}
\label{P}
P(\omega, {\bm k})=1-\frac{2\omega_{pe}^2}{\omega^2}\,W\left(\omega, k_z\right).
\end{eqnarray}
The function $W$ in this expression is given by 
\begin{eqnarray}
W=-\frac{\omega^2}{k_z}\int\limits_{-c}^{c}\frac{1}{\omega-k_zv_z +i\nu}\,\frac{d{\tilde f}}{dv_z}\,dv_z,
\label{W}
\end{eqnarray}
where $\nu\to +0$ is needed to describe collisionless Landau damping. Let us first discuss the limit of large parallel phase velocity, $\omega \gg k_z v_{th}$, where $v_{th}$ is the characteristic (e.g., thermal) speed of the particle distribution. Obviously, this limit can also describe the case of cold non-relativistic plasma, when plasma temperature is negligibly small. In this limit, we can neglect the imaginary part and integrate Eq.~(\ref{W}) by parts: 
\begin{eqnarray}
W=\omega^2\int\limits_{-c}^{c}\frac{\tilde f}{\left(\omega -k_zv_z\right)^2}\, dv_z\approx\int\limits_{-c}^{c} {\tilde f}\,dv_z =\int\limits_{-\infty}^{\infty}\left(1-\frac{v_z^2}{c^2}\right)^{3/2}{\tilde f}\, du_z \equiv \left\langle \frac{1}{\gamma^3} \right\rangle.
\end{eqnarray}
One can then define the plasma frequency for relativistic pair plasma as follows:
\begin{eqnarray}
\omega_p^2\equiv 2\omega_{pe}^2\left\langle \frac{1}{\gamma^3} \right\rangle,
\label{rel_omega}
\end{eqnarray}
which provides the relativistic generalization of the non-relativistic expression. The function $P$ now has the form
\begin{eqnarray}
P=1-\frac{\omega_p^2}{\omega^2}.
\end{eqnarray}
Substituting this function into Eq.~(\ref{disp}), we obtain the dispersion relation: 
\begin{eqnarray}
\label{omega_pm}
\omega^{2}=\frac{\left(\omega_{p}^{2}+k^{2}c^{2}\right)\pm\sqrt{\left(\omega_{p}^{2}+k^{2}c^{2}\right)^{2}-4\omega_{p}^{2}k_{z}^{2}c^{2}}}{2}.
\end{eqnarray}
Here, the positive sign in front of the square root corresponds to the {\it ordinary} mode, while the negative sign to the {{\it Alfv\'en} mode, which transforms into the inertial-Alfv\'en mode at $k>\omega_p/c$}. In the case of cold plasma, both solutions are allowed. However, in our case of relativistic plasma temperature, the latter solution is not applicable as it would correspond to the parallel phase speed smaller than the thermal speed. We, therefore, analyze only the expression corresponding to the ``$+$''~sign in~Eq.~(\ref{omega_pm}).

In the long-wave limit, $kc\ll \omega_p$, this expression gives $\omega = \omega_p$, and the corresponding electric field is polarized along the background magnetic field, ${\bm E}=(0,0,E_z)$. In the opposite limit, $kc\gg \omega_p$, we get $\omega = kc$. One can check that in this case, the electric field lies in the $x-z$ plane and it is normal to the wave vector ${\bm k}$. In the limit of quasi-perpendicular wave propagation, $k_\perp\gg k_z$, the dispersion relation for the ordinary mode simplifies to
\begin{eqnarray}
\omega^2=\omega_p^2+k^2c^2
\label{omega_o}
\end{eqnarray}
for any value of the wavenumber, with the electric polarization being nearly aligned with the background magnetic field, 
\begin{eqnarray}
{\bm E}_{\mbox{{\tiny O}}}\approx (0,0, E_z).
\end{eqnarray}

We now consider the limit of low phase velocities, $\omega\sim k_z v_{th}$, which will give us the dispersion relation for the Alfv\'en mode. In this limit, the dispersion relation depends on the details of the particle distribution function. As was discussed previously, magnetically dominated turbulence with ${{\tilde \sigma}_0 \gg 1}$ leads to ultrarelativistic particle heating, with $v_{th}\approx c$. We will therefore assume relativistic distribution functions in our consideration. For instance, one can consider the one-dimensional equilibrium Maxwell-J\"uttner distribution, 
\begin{eqnarray}
{\tilde f}(u_z)=\frac{1}{2K_1(1/\theta)c}\exp\left(-{\gamma}/{\theta}\right),
\end{eqnarray}
where $\gamma=1/\sqrt{1-v_z^2/c^2}$, $K_1$ is the modified Bessel function of the second kind, and $\theta=k_BT/mc^2$. For ultrarelativistic particle temperatures, $\theta \gg 1$, one can replace $K_1(1/\theta)\approx\theta$. In what follows, we will also need to know the enthalpy density $w$ corresponding to this distribution. For ultrarelativistic one-dimensional gas, the internal energy density and pressure are related as $P_\|=u$, and we derive the normalized enthalpy density: 
\begin{eqnarray}
\label{w1D}
w=(u + P_\|)/n_0m_ec^2=2\theta. 
\end{eqnarray}

We now assume without loss of generality that $k_z>0$, and rewrite the expression for the $W$ function (\ref{W}) as follows:
\begin{eqnarray}
W=-\frac{\omega^2}{k^2_zc}\int\limits_{-c}^{c}\frac{1}{\left(1-{v_z}/{c}\right)-\left(1-{\omega}/{k_zc}\right) +i\nu}\,\frac{d{\tilde f}}{dv_z}\,dv_z.
\label{W2}
\end{eqnarray}
The distribution function declines fast when particle energy exceeds the thermal energy, that is when $\gamma > \theta$, or equivalently $1-v_z/c < 1/(2\theta^2)$, where we have approximated $1/\gamma^2\approx 2(1-v_z/c)$.  The integral is thus dominated by the velocity values satisfying $1-v_z/c\sim 1/(2\theta^2)$. Therefore, the behavior of the ultrarelativistic $W$-function depends on whether $1-\omega/k_zc$ is greater or smaller than the small parameter $1/(2\theta^2)$. It is easy to see that this is just the condition that compares the parallel phase velocity of the waves, $\omega/k_z$, with the velocity associated with the thermal motion of the particles, $v_{th}= c\sqrt{1-1/\theta^2}$. These two asymptotic limits should be considered separately; we refer the reader to \citet{godfrey1975,gedalin1998} for a detailed analysis. For our consideration of the ultrarelativistic Alfv\'en mode with $k_z\ll k_\perp$, the essential limit is 
\begin{eqnarray}
\label{limit1}
\left| 1-\frac{\omega}{k_zc}\right| \ll \frac{1}{2\theta^2}.
\end{eqnarray}
In this limit, the imaginary part of the distribution function is negligible. Moreover, in this case, one can neglect $1-\omega/k_zc$ with respect to $1-v_z/c$ in the denominator. The asymptotic expression for the $W$-function in this limit can then be found from Eq.~(\ref{W2}) where we integrate by parts, 
\begin{eqnarray}
W \sim \frac{\omega^2}{k^2_zc^2}\int\limits_{-c}^{c}\frac{1}{(1-{v_z}/{c})^2}\,{{\tilde f}}\,dv_z
\sim \frac{\omega^2}{k^2_zc^2}\int\limits_{0}^{c}4\gamma^4\,{{\tilde f}}\,dv_z
= \frac{\omega^2}{k^2_zc^2}\int\limits_{0}^{\infty}4\gamma\,{{\tilde f}}\,du_z
= \frac{\omega^2}{k^2_zc^2}\,2\left\langle \gamma \right\rangle.
\end{eqnarray}
The average value of $\gamma$ depends on the distribution function. For the considered ultrarelativistic Maxwell-J\"uttner distribution, one gets $\langle\gamma \rangle= \theta$. Substituting this into Eq.~(\ref{P}), one obtains
\begin{eqnarray}
P=1-\frac{4\theta\omega_{pe}^{2}}{k_z^2 c^2}\approx -\frac{4\theta\omega_{pe}^{2}}{k_z^2 c^2}.
\end{eqnarray}
Eq.~(\ref{disp}) then leads to the dispersion relation of the ultrarelativistic Alfv\'en mode
\begin{eqnarray}
\label{A}
\omega^2=k_z^2c^2\left(1-\frac{k_\perp^2 c^2}{4\omega^2_{pe}\theta}  \right)=k_z^2c^2\left(1-\frac{k_\perp^2 d_{rel}^2}{w^2}  \right),
\end{eqnarray}
where we have defined the relativistic inertial scale of a pair plasma as
\begin{eqnarray}
d_{rel}^2 = c^2\frac{\theta}{\omega_{pe}^2}= c^2\frac{w}{2\omega_{pe}^2}.
\end{eqnarray}
Recalling that our derivation is valid under the assumption that $1-\omega^2/k_z^2c^2\ll 1/\theta^2$, we see that the Alfv\'en dispersion relation (\ref{A}) holds up to the scales $k_\perp^2d_{rel}^2\sim 1$. At perpendicular scales comparable to the relativistic inertial scale, the thermal effects become essential and Landau damping becomes strong.  The polarization of the Alfv\'en mode is given by 
\begin{eqnarray}
{\bm E}_{\mbox{\tiny A}}\approx (E_x, 0, 0),    
\end{eqnarray}
so that this mode is nearly potential. 


Finally, we discuss the Alfv\'en wave for the case of the so-called ``waterbag" distribution, which has cut-offs in the momentum space. Such functions were used to study relativistic beams in a plasma \cite[e.g.,][]{roberts1967,davidson2004},  to model pair plasma distributions in pulsar magnetospheres \cite[][]{arons1986}, and analyzed in detail in \cite[][]{gedalin1998}. We will however modify the ``waterbag" distribution by smoothing out its sharp edges. This can be done in many different ways by introducing various regularizations whose particular forms are not relevant for our consideration. We may, for example, adopt the following model function
\begin{eqnarray}
{\tilde f}(u_z)=A\frac{1}{e^\frac{\gamma-\gamma_m}{\theta} +1}.   
\end{eqnarray}
In the limit $\theta/\gamma_m\to 0$, such a distribution approaches (up to the normalization constant $A$) the Heaviside step function $H(\gamma_m-\gamma)$. We however assume that $\theta$ is small but nonzero, $0< \theta/\gamma_m\ll 1$, in which case the sharp boundary of the step function is smoothed over a narrow region $\Delta \gamma\sim \theta$. In the ultrarelativistic case, $\gamma_m\gg 1$, the normalization constant is $A=1/(2c\gamma_m)$ and the enthalpy density corresponding to such a distribution is given by $w=\gamma_m$. The derivative of the distribution function is then a $\theta$-broadened delta function, $c\,d{\tilde f}/du_z=A \,\delta(\gamma-\gamma_m)$, and we can integrate Eq.~(\ref{W}) to obtain \cite[][]{gedalin1998}:
\begin{eqnarray}
W= \frac{1}{\gamma_m}\frac{\omega^2}{k_z^2v_m^2}\frac{1}{(\omega^2/k_z^2v_m^2-1)},   
\end{eqnarray}
where $1/\gamma_m^2=1-v_m^2/c^2$. The resulting dispersion relation for the Alfv\'en mode is then
\begin{eqnarray}
\label{A_wb}
\omega^2=k_z^2c^2\frac{1+{k_\perp^2v_m^2\gamma_m}/{2\omega_{pe}^2}}{1+{k_\perp^2c^2\gamma_m}/{2\omega_{pe}^2}}= k_z^2c^2\frac{1+\frac{v_m^2}{c^2}k_\perp^2d_{rel}^2}{1+{k_\perp^2d_{rel}^2}}, \end{eqnarray}
where we use the relativistic inertial scale  
\begin{eqnarray}
d_{rel}^2=c^2\frac{\gamma_m}{2\omega_{pe}^2}= c^2\frac{ w}{2\omega_{pe}^2}.    
\end{eqnarray}
These expressions agree with our hydrodynamic result (\ref{iaw}). For perpendicular scales larger than the inertial scale, $k_\perp^2d_{rel}^2\ll 1$, we obtain the familiar ultrarelativistic Alfv\'en mode, cf.~(\ref{A}),
\begin{eqnarray}
\label{A2}
\omega^2\approx k_z^2c^2\left(1-\frac{k_\perp^2d_{rel}^2}{\gamma_m^2} \right)=k_z^2c^2\left(1-\frac{k_\perp^2d_{rel}^2}{w^2} \right),
\end{eqnarray}
while in the opposite, ``kinetic" range $k_\perp^2d_e^2\gg 1$, the dispersion relation changes to
\begin{eqnarray}
\omega^2\approx k_z^2v_m^2\left(1+ \frac{1}{k_\perp^2d_{rel}^2 w^2}\right).
\label{omega_wb}
\end{eqnarray}
The parallel phase velocity of this mode exceeds~$v_m$. The Landau damping is weak if this phase velocity is not too close to $v_m$, that is if their difference is larger than the boundary broadening, $1/\sqrt{1-\omega^2/k_z^2c^2}-\gamma_m>\theta$. Substituting here expression (\ref{omega_wb}), we derive the condition for such a mode to exist: 
\begin{eqnarray}
k_\perp^2d_{rel}^2 < \frac{\gamma_m}{2\theta}.     
\end{eqnarray}
We note the analogy of the ``waterbag" distribution with the Fermi distribution in a degenerate gas. The acoustic-type mode (\ref{omega_wb}) appearing at kinetic scales in this case is then analogous to the so-called ``zero sound," $\omega\approx kv_F$, existing in a degenerate plasma where the particles have a Fermi distribution with a cutoff at the Fermi speed~$v_F$. We also point out that different kinetic particle distributions lead to similar results for the relativistic Alfv\'en mode when expressed in terms of the fluid parameters $d_{rel}$ and $w$; this can be seen from comparing Eqs.~(\ref{A}) and (\ref{A2}).






\bibliography{references}

\end{document}